# Do altmetric scores reflect article quality? Evidence from the UK Research Excellence Framework 2021

Mike Thelwall
Statistical Cybermetrics and Research Evaluation Group, University of Wolverhampton, UK.
https://orcid.org/0000-0001-6065-205X m.thelwall@wlv.ac.uk

Kayvan Kousha
Statistical Cybermetrics and Research Evaluation Group, University of Wolverhampton, UK.
https://orcid.org/0000-0003-4827-971X k.kousha@wlv.ac.uk

Mahshid Abdoli
Statistical Cybermetrics and Research Evaluation Group, University of Wolverhampton, UK.
https://orcid.org/0000-0001-9251-5391 m.abdoli@wlv.ac.uk

Emma Stuart
Statistical Cybermetrics and Research Evaluation Group, University of Wolverhampton, UK.
https://orcid.org/0000-0003-4807-7659 emma.stuart@wlv.ac.uk

Meiko Makita
Statistical Cybermetrics and Research Evaluation Group, University of Wolverhampton, UK.
https://orcid.org/0000-0002-2284-0161 meikomakita@wlv.ac.uk

Paul Wilson
Statistical Cybermetrics and Research Evaluation Group, University of Wolverhampton, UK.
https://orcid.org/0000-0002-1265-543X pauljwilson@wlv.ac.uk

Jonathan Levitt
Statistical Cybermetrics and Research Evaluation Group, University of Wolverhampton, UK.
https://orcid.org/0000-0002-4386-3813 j.m.levitt@wlv.ac.uk

Altmetrics are web-based quantitative impact or attention indicators for academic articles that have been proposed to supplement citation counts. This article reports the first assessment of the extent to which mature altmetrics from Altmetric.com and Mendeley associate with journal article quality. It exploits expert norm-referenced peer review scores from the UK Research Excellence Framework 2021 for 67,030+ journal articles in all fields 2014-17/18, split into 34 Units of Assessment (UoAs). The results show that altmetrics are better indicators of research quality than previously thought, although not as good as raw and field normalised Scopus citation counts. Surprisingly, field normalising citation counts can reduce their strength as a quality indicator for articles in a single field. For most UoAs, Mendeley reader counts are the best, tweet counts are also a relatively strong indicator in many fields, and Facebook, blogs and news citations are moderately strong indicators in some UoAs, at least in the UK. In general, altmetrics are the strongest indicators of research quality in the health and physical sciences and weakest in the arts and humanities. The Altmetric Attention Score, although hybrid, is almost as good as Mendeley reader counts as a quality indicator and reflects more non-scholarly impacts.

# Introduction

Altmetrics are quantitative indicators for research outputs that are not based on traditional citations from journal articles but are usually derived from web sources. They are widely found in publisher websites, usually sourced from the Altmetric.com or PlumX data providers, although CrossRef also provides relevant data (Ortega, 2018). Altmetrics can also be found in the free scholarly search engine Dimensions. Most academics seem to be aware of some of them (Aung et al., 2019), testifying to their importance within the scholarly communication ecosystem. Evidence about the information contained in altmetrics is needed for them to be interpreted effectively, however (Haustein et al., 2014c; Sud & Thelwall, 2014). This is complicated by a lack of quality control for most and the potential for many of them to be gamed or infiltrated by irrelevant data (Roemer & Borchardt, 2015; Wouters & Costas, 2012). Nevertheless, they may have value for formative evaluation, if used carefully (Bar-Ilan et al., 2018).

The rationale for citation-based impact indicators is that citations can reflect the cited document influencing the citing document, so citation counts partly reflect scholarly influence or impact. Although perfunctory citations also occur, it is still reasonable to use citation counts as scholarly impact indicators if relatively trivial citations can be ignored as "noise" in the system (Moed, 2006). In contrast, the various altmetrics have been hypothesised to reflect different dimensions of attention or impact, and especially societal impact (Priem et al., 2011; Kousha, 2019). Most also have the advantage of appearing before citation counts, giving earlier evidence of interest or impact. There is substantial evidence that one altmetric, Mendeley reader counts, is a scholarly impact indicator and a partial educational impact indicator for journal articles primarily because Mendeley reader counts correlate moderately or strongly with citation counts for articles in most academic fields (Thelwall & Sud, 2016) and can be used as early scholarly impact indicators (Zahedi et al., 2017). Nevertheless, the value and best interpretations of all other altmetrics are uncertain. Tweeter counts, for example, although having moderate correlations with citation counts in some fields (Costas et al., 2015; Haustein et al., 2014a), seem to reflect academic interest and author/publisher dissemination activities in many fields rather than the initially hypothesised public interest (e.g., Lemke et al., 2022), despite most Twitter users being non-academics. Biomedical research might be an exception because this research is widely tweeted by the public (Mohammadi et al., 2018; see also: Haustein et al., 2014b).

The reason for the ongoing uncertainty about how to interpret altmetrics is a lack of relevant data. Although there are many ways to partially evaluate altmetrics (Sud & Thelwall, 2014), there is no large-scale systematic evidence of the attention given to, or societal impact of, academic research. Thus, there is no direct way to check which altmetrics can reasonably be claimed to be indicators of these, or in which fields. Given this absence, the most common approach has been to correlate altmetric scores with citation counts, as an indicator of scholarly impact, on the basis that positive correlations would at least indicate that altmetric scores are non-random and scholarly-related to some extent. This is almost a paradox since the value of most altmetrics would be in being different from citation counts, but an overlap could nevertheless be expected for any scholarly-related indicator (Thelwall, 2016). Other methods previously used have included content analyses of individual sources (e.g., tweets:

Holmberg & Thelwall, 2014), and predicting future citation counts from early altmetric scores (Akella et al., 2021; Thelwall & Nevill, 2018).

Since citation counts are not direct measures of scholarly impact, a better way to evaluate altmetrics would be to correlate them against peer review quality scores for journal articles. This is more direct and may reveal altmetrics that reflect dimensions of quality not well captured by citations. This is plausible since significance (i.e., impact, whether scholarly, societal or other) is one of the three core components of quality, with the other two being rigour and originality (Langfeldt et al., 2020). One non peer reviewed publication has previously done this. It correlated a range of altmetrics, including Mendeley reader counts and tweet counts from Altmetric.com, with Research Excellence Framework (REF) expert peer review quality scores for 19,580 journal articles from 2008 in 36 field-based Units of Assessment (UoAs). It found only relatively low correlations with quality scores, with the highest in Clinical Medicine (rho=0.441) and Biological Sciences (rho=0.363), (HEFCE, 2015), undermining previous claims for the usefulness of altmetrics. A limitation of the analysis was that Altmetric.com started in 2011, so its data for 2008 may have been incomplete. Moreover, given the relatively low numbers of articles in some UoA, some results may have been imprecise.

The current article updates the REF2014 technical report with more current REF2021 data on the basis that altmetrics have matured over time and Altmetric.com data may be more comprehensive after 2011. Data maturation is likely because the only year previously analysed, 2008, precedes Altmetric.com's foundation in 2011 and immediately follows Mendeley's creation in 2007. The primary research question is to assess the overall value of altmetrics. The second relates to the Altmetric Attention Score because, although it is a hybrid indicator and therefore not useful to study from a theoretical perspective, it is Altmetric.com's flagship indicator and one of the best known altmetrics. It is important to understand whether it has value from the user perspective, given that many researchers must notice and perhaps consult it. The final research question benchmarks against citation counts, as the most widely used research impact indicator.

- RQ1: How useful are Altmetric.com altmetrics as indicators of research quality in all fields?
- RQ2: How useful is the Altmetric Attention Score as an indicator of research quality in all fields?
- RQ3: How do altmetrics compare to raw and field normalised citation counts as indicators of research quality in all fields?

# Methods

## Data

Provisional REF2021 scores from March 2022 for 148,977 journal articles from 2014 to 2020 were supplied by the REF team as part of an unrelated project. This includes many duplicate articles that were scored separately because they were supplied by different authors. For security reasons, University of Wolverhampton submissions were excluded. Each score had been agreed by two subject experts, usually senior researchers, from one of the 34 UoAs and agreed at the UoA level, with norm referencing within each UoA. Thus, the scores are carefully calibrated expert judgements. Each output was scored as 0 unclassified, 1* recognised nationally, 2* recognised internationally, 3* internationally excellent, or 4* world-leading in terms of originality, significance, and rigour (REF2021, 2021). Since the score 0 could be allocated to a very weak article or a stronger article with a technical noncompliance, the few articles with score 0 were removed. Articles without Digital Object Identifiers (DOIs) were also excluded (difficult to match Altmetric.com data), as were articles not in Scopus (needed for the citation RQ). Duplicate articles within a UoA were removed, allocating the remaining article the median score of all copies (randomly rounding up or down when the median was a x.5 fraction). Articles after 2018 were excluded because of insufficient time to attract a stable number of citations (Wang, 2013). Exact numbers for each field and year are in the supplementary materials but there were low numbers for the arts and humanities UoAs.

Altmetric scores were obtained for each article 2014-18 by querying its DOI with the Altmetric API (Robinson-García et al., 2014) in Webometric Analyst (https://lexiurl.wlv.ac.uk/) for Altmetric's public record. Altmetric was chosen in preference to PlumX for its free API that allowed batch downloading of records for all articles. Articles without a record in Altmetric.com were excluded from the 2014-18 data. It would also have been reasonable to assume that such articles had altmetric scores of 0, but some may also have had Altmetric records with a different DOI (either configured differently or for a different version of the article, such as a conference paper, preprint, or update). For example, UoA 11 Computer Science and Informatics had a large minority of articles without DOI matches in Altmetric.com. Investigations of these articles found that they sometimes had Altmetric scores associated with an ArXiv DOI for the preprint of the article. Altmetric.com presumably knew the official DOI but used the preprint DOI as the primary source for API queries and the online record. Thus, assuming articles with DOIs without Altmetric.com API query matches would have altmetric scores of 0 would sometimes be false.

Since Altmetric's Mendeley data may not be systematically updated for all DOIs, Mendeley records were captured directly from Mendeley using its API, again in Webometric Analyst for each article 2014-17 (see below for the rationale for excluding 2018). Finally, for comparison, Scopus citation counts from January 2021 for each article 2014-17 were also added by DOI. Sample sizes are in Table 1.

Table 1. Numbers of journal articles used in the analyses. Main Panel data excludes duplicates within UoAs and the 2014-18 data excludes articles without Altmetric.com records.

| Unit of Assessment or Main Panel | 2014-17 articles | 2014-18 articles |
|---|---:|---:|
| 1: Clinical Medicine | 5735 | 7068 |
| 2: Public Health, Health Services and Primary Care | 2259 | 2824 |
| 3: Allied Health Professions, Dentistry, Nursing and Pharmacy | 5517 | 6216 |
| 4: Psychology, Psychiatry and Neuroscience | 4662 | 5499 |
| 5: Biological Sciences | 3744 | 4610 |
| 6: Agriculture, Food and Veterinary Sciences | 1747 | 1972 |
| 7: Earth Systems and Environmental Sciences | 2205 | 2522 |
| 8: Chemistry | 2063 | 2162 |
| 9: Physics | 3090 | 3333 |
| 10: Mathematical Sciences | 2922 | 2037 |
| 11: Computer Science and Informatics | 2504 | 1749 |
| 12: Engineering | 10036 | 6236 |
| 13: Architecture, Built Environment and Planning | 1280 | 1038 |
| 14: Geography and Environmental Studies | 1766 | 2099 |
| 15: Archaeology | 279 | 331 |
| 16: Economics and Econometrics | 913 | 760 |
| 17: Business and Management Studies | 5868 | 4847 |
| 18: Law | 852 | 877 |
| 19: Politics and International Studies | 1224 | 1417 |
| 20: Social Work and Social Policy | 1570 | 1834 |
| 21: Sociology | 720 | 848 |
| 22: Anthropology and Development Studies | 469 | 526 |
| 23: Education | 1592 | 1666 |
| 24: Sport and Exercise Sciences, Leisure and Tourism | 1368 | 1625 |
| 25: Area Studies | 223 | 241 |
| 26: Modern Languages and Linguistics | 476 | 366 |
| 27: English Language and Literature | 367 | 261 |
| 28: History | 535 | 535 |
| 29: Classics | 50 | 28 |
| 30: Philosophy | 368 | 348 |
| 31: Theology and Religious Studies | 83 | 59 |
| 32: Art and Design: History, Practice and Theory | 522 | 414 |
| 33: Music, Drama, Dance, Performing Arts, Film and Screen Studies | 291 | 217 |
| 34: Communication, Cultural & Media Studies, Library & Information Man. | 436 | 465 |
| Main Panel A (UoAs 1-6) | 21327 | 25239 |
| Main Panel B (UoAs 7-12) | 22145 | 17353 |
| Main Panel C (UoAs 13-24) | 17494 | 17392 |
| Main Panel D (UoAs 25-34) | 3324 | 2905 |
| **Total (UoAs 1-34)** | **67736** | **67030** |

The Scopus citation counts were converted into Normalised Log-transformed Citation Scores (NLCS) to give a theoretically better citation-based indicator (Thelwall, 2017). Field normalisation of citations in scientometrics is common because fields citation rates vary between fields and the normalisation process factors this out (Waltman et al., 2011). NLCS values were obtained by first log-transforming all citation scores with log(1+x) to reduce skewing, then averaging the log-transformed values separately for each Scopus narrow field and year. Each article NLCS was then calculated as its log-transformed citation count divided by the average for its field and year. An article in multiple fields would instead have its log(1+x) divided by the average over all relevant fields. The fields used for this were Scopus narrow fields (Scopus, 2022), which are approximately 325 (depending on year) different categories. An NLCS value of 1 indicates world average citation impact, irrespective of the field and year of the article and the scores are comparable between years and fields.

### Analysis

Spearman correlations were used to assess the strength of association between REF2021 expert peer review scores and altmetric scores for all research questions. Spearman correlations were used instead of Pearson correlations since citation and altmetric data can be highly skewed (Thelwall & Wilson, 2016). Whilst correlation does not show cause-and-effect, positive Spearman correlations suggests that articles with a higher altmetric or citation count tend to have higher REF2021 quality scores. Correlations were calculated separately for each year to reduce the influence of time on the results. For citations counts, at least a three-year window is usually adequate to get reliable results (Wang, 2013), so only articles from 2014-17 were used for the citation data and, since it is compared to the Mendeley API data, the same was applied to the latter. To keep the three-year window, articles from 2014-18 were used for the Altmetric.com data collected in 2022. For ease of reporting masses of results, the median correlation across all relevant years was reported, but results for individual years are in the supplementary material.

## Results

### Correlations

The correlation results are shown here for the altmetrics supplied by the Altmetric API except those for which the median correlations were below 0.1 for all UoAs: Pinners, Questions, GPlus. The remaining results are displayed in themed batches because there is too much data to fit on one graph. A common scale is used for ease of comparison between graphs.

Both Scopus citations and Mendeley readers have similar levels of correlation with REF2021 provisional quality scores in most UoAs, but the Scopus correlations are higher in all except two (17, 34). Mendeley readers seem to be particularly weak in the humanities. This might be because Mendeley is a reference manager and humanities reference styles are often based on discussions in footnotes rather than standard format references. Thus, Mendeley is less useful to such scholars and its records may be sparser (Thelwall, 2019). The relatively low correlations for Mendeley in Mathematical Sciences and Computer Science and Informatics are presumably due to the LaTeX document formatting language commonly used in these areas (also in parts of Physics), for which Mendeley would be less use. The results confirm that citations and Mendeley readers have the most information value in medicine, health, physical sciences, moderate value in mathematics, engineering, and social sciences, and little in the arts and humanities.

Comparing the article quality correlations for the NLCS field normalised citation counts with those for the raw Scopus citation counts, it is surprising that raw citation counts are better indicators of research quality in over two thirds of UoAs (with nine exceptions: 10, 16,

18, 21, 25, 28, 31, 32, 33). This is surprising because field normalised indicators are designed to be fairer than raw citation counts by taking into account the publication field, so an article does not have an advantage for being published in a high citation speciality. In this case the correlations are calculated within field-based UoAs, so field normalisation should make little difference. Nevertheless, articles submitted to UoAs by their UK authors can be interdisciplinary or submitted to out-of-field UoA (e.g., because the author is a statistician in a medical department), which the NLCS normalisation process should help with. Mostly lower correlations for NLCS suggest that the field normalisation process is flawed. This is plausible since Scopus categorises articles by journal, but article-level classifications more closely align with underlying topics (Klavans & Boyack, 2017). Thus, the results suggest that field normalisation, at least based on articles classified by journal, is usually counterproductive when analysing articles from a single broad field and year.

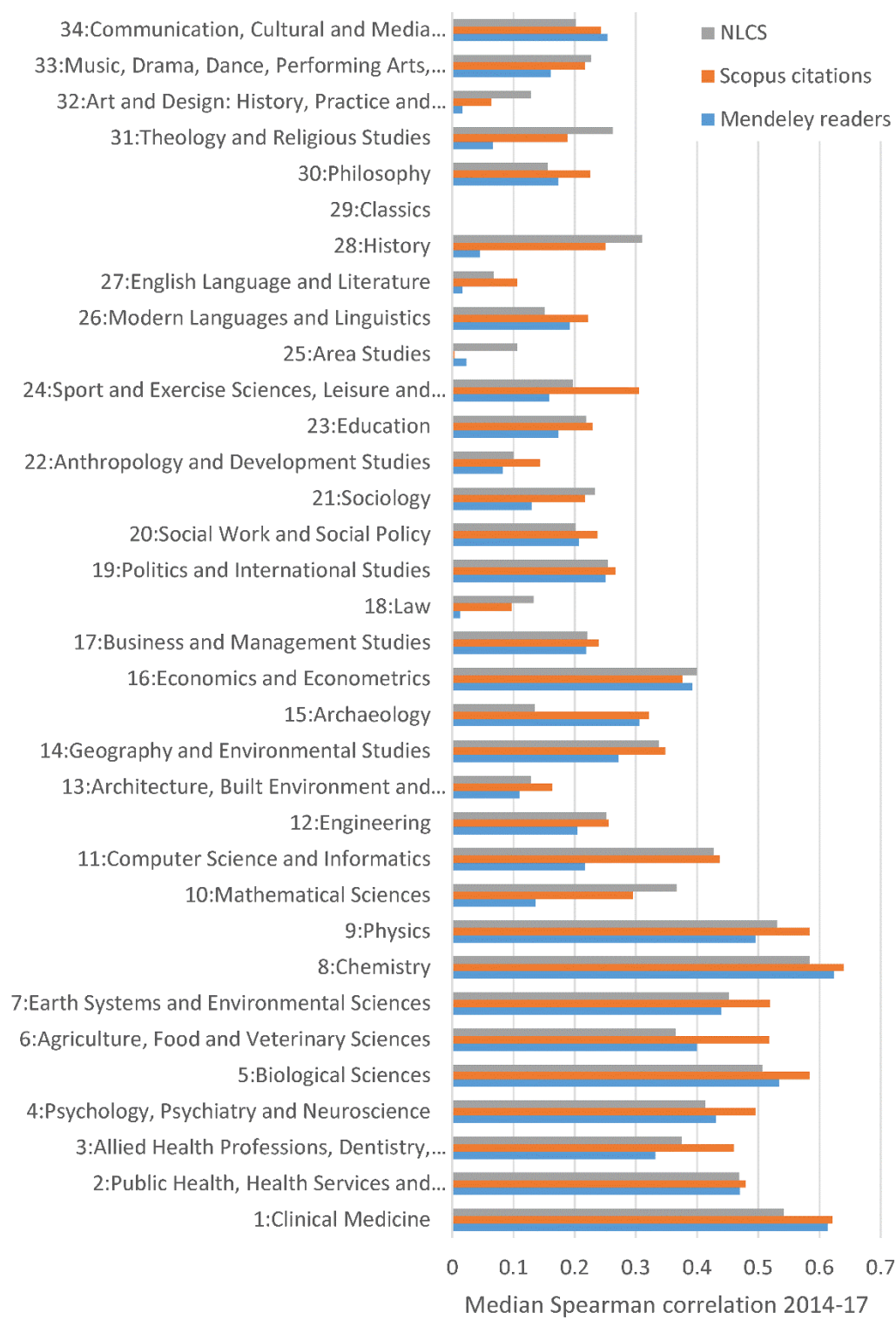

Figure 1. Scopus citations (count and NLCS) and Mendeley readers (from Mendeley API): Spearman correlations with provisional REF2021 scores, calculated separately for each UoA and year, with the median across years reported. UoA 29 results have been removed for single figure sample sizes.

The Altmetric.com data for Mendeley gives similar correlations to the Mendeley API data (Figure 2). Altmetric.com also counts readers from CiteULike, but this is used by many fewer people, which is the likely cause of lower correlations with REF2021 scores in all UoAs except

Area Studies. Combining the CiteULike with the Mendeley counts to give Total Readers does not tend to improve on the Mendeley reader count correlation, so Mendeley readers alone are sufficient. The Altmetric Attention Score (AAS) is a "weighted count of all of the attention a research output has received" (excluding Mendeley readers), based on volume, source type and authorship (Altmetric, 2022b). Although Mendeley readers are better indicators of research quality in most UoAs and substantially better in some medical and physical sciences fields, AAS is slightly better for a third, mostly from the arts, humanities, and social sciences: UoAs 6, 7, 14, 20, 22, 23, 25, 28, 30, 31, 34.

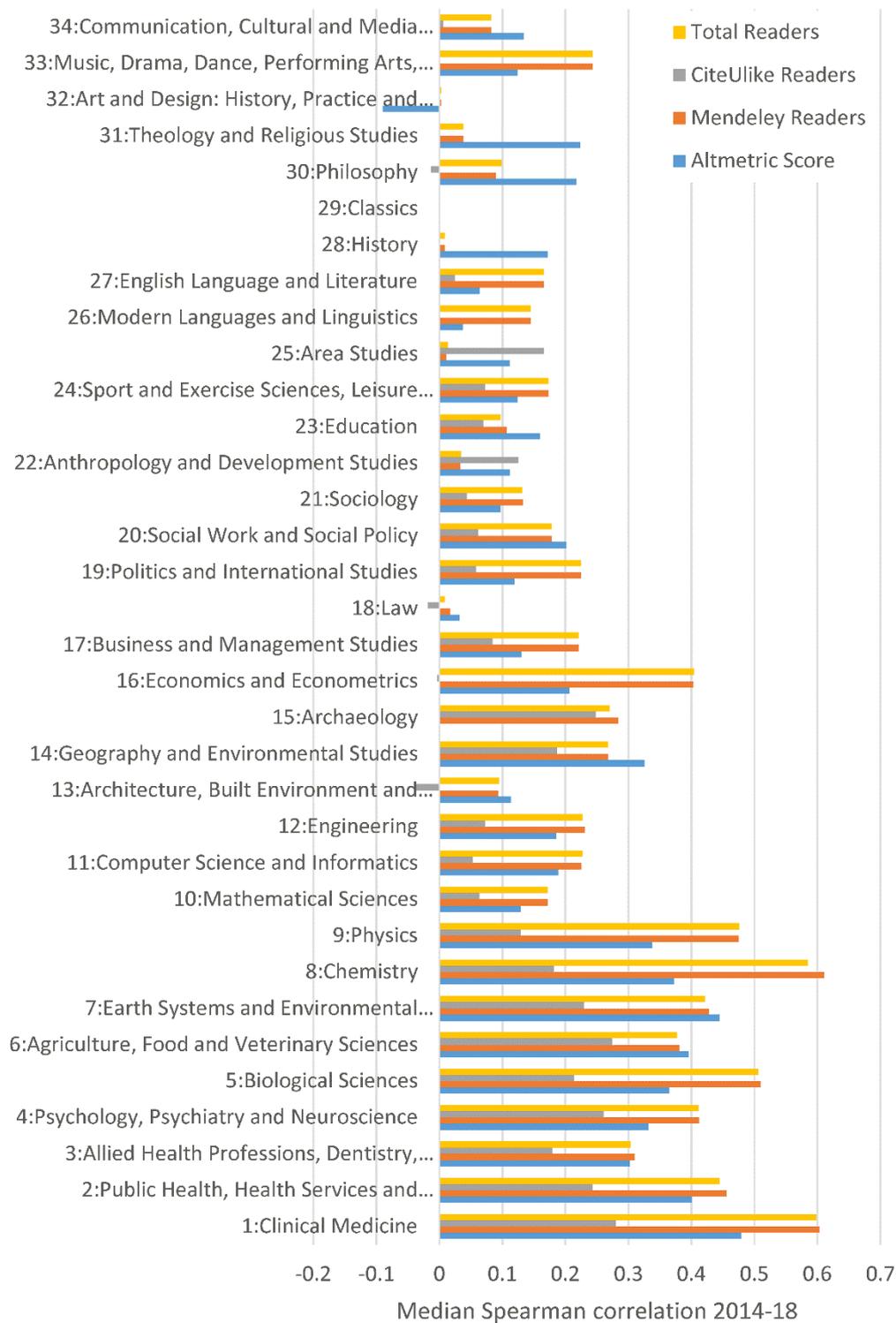

Figure 2. Altmetric Scores and online reference manager readers: Spearman correlations with provisional REF2021 scores, calculated separately for each UoA and year, with the median across years reported. UoA 29 results have been removed for single figure sample sizes.

Of the news related sources, Tweeters (the number of Twitter users tweeting an article URL, apparently from a list of academic Tweeters: Altmetric, 2022a) seems to be the best indicator of research quality (Figure 3). Nevertheless, Blog and news citations (both from curated lists of sources: Altmetric, 2022a) also have moderate strength as research quality indicators in many UoAs. Facebook Wall links (from a curated list of walls: Altmetric, 2022a) are the weakest, presumably due to smaller numbers of academically-relevant walls curated. Twitter is weaker than Altmetric's Mendeley readers as a research quality indicator in over three quarters of UoAs. The exceptions are mostly in the social sciences, arts, and humanities: UoAs 6, 13, 14, 18, 22, 25, 28, 30, 34.

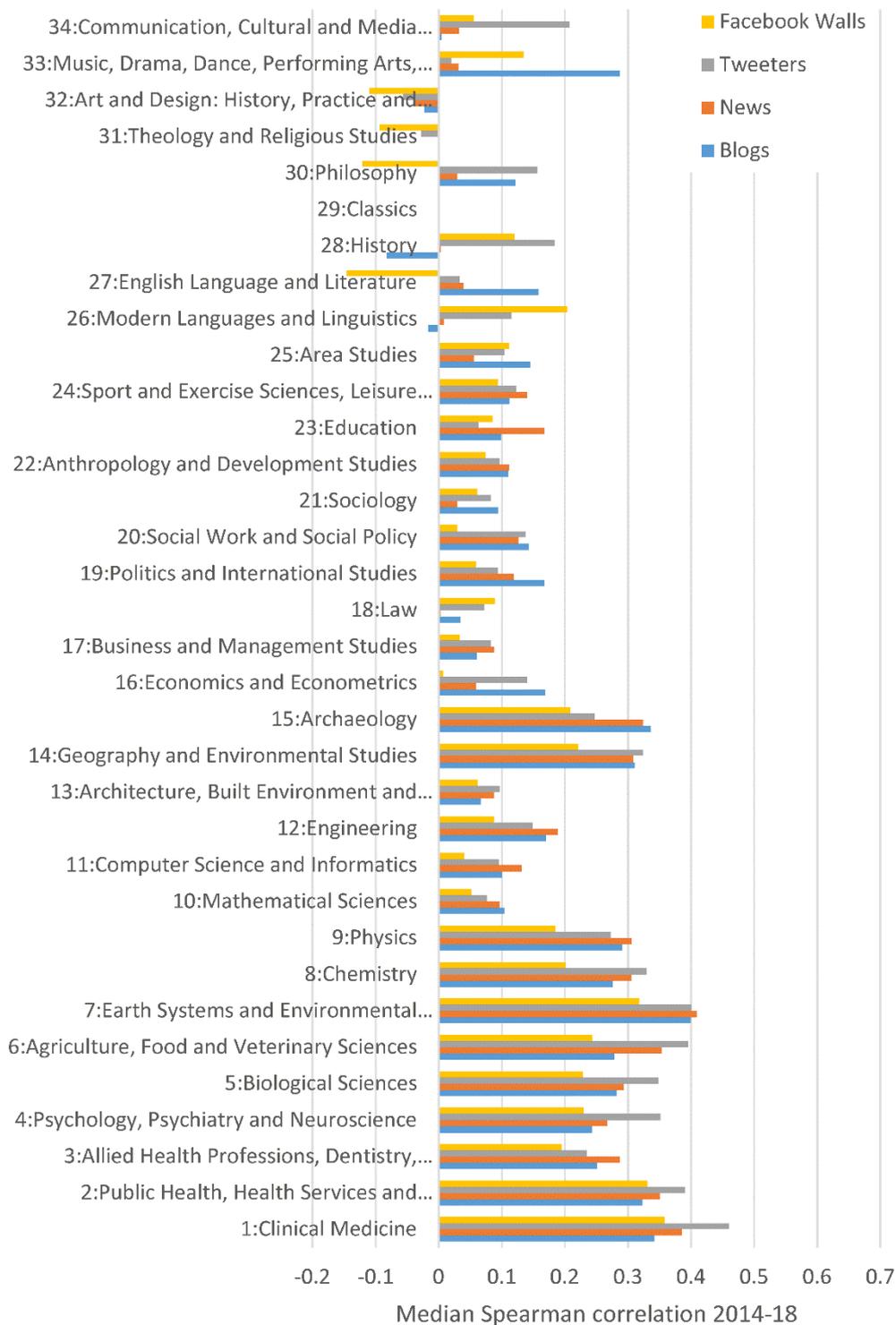

Figure 3. Social network and news sites: Spearman correlations with provisional REF2021 scores, calculated separately for each UoA and year, with the median across years reported. UoA 29 results have been removed for single figure sample sizes.

Reddit mentions, Wikipedia citations and research highlight reviews ("Recommendations of individual research outputs from Faculty Opinions": Altmetric, 2022a) are all weak indicators of research quality in all fields, presumably for their scarcity. Nevertheless, Wikipedia citations have a moderate correlation with research quality in Archaeology and perform well compared to Mendeley Readers and Tweeters in some arts and humanities subjects (Figure 4).

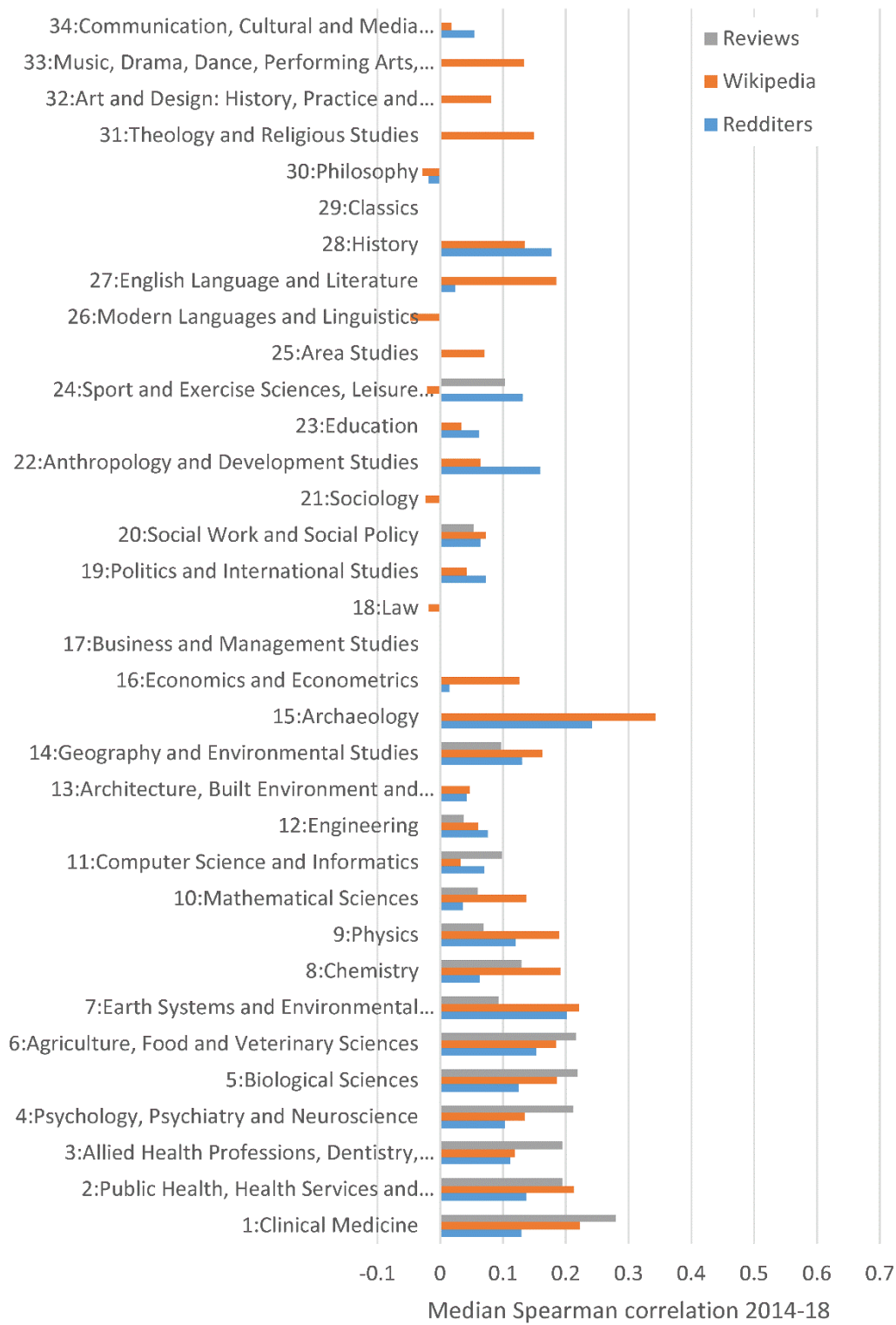

Figure 4. Wikipedia, Reddit and facultyopinions.com research highlight reviews: Spearman correlations with provisional REF2021 scores, calculated separately for each UoA and year, with the median across years reported. UoA 29 results have been removed for single figure sample sizes.

## Discussion

The results are limited by the restriction to the UK and by the articles analysed being self-selected by academics to be their best work from 2014-20. Thus, the relatively low proportions of weaker research in the sets used for correlation probably reduces the strength of the correlations. In particular, there are few low quality 1* articles and the absence of a substantial proportion of low quality articles that may well score of 0 on all indicators would reduce all correlations. Conversely, since the UK is a heavy user of social media, including Mendeley, Twitter and Facebook, it is likely that similar correlations would be lower for most other countries. The value of altmetrics may also change over time as the demographics of their users shift. For example, the desktop version of Mendeley started to be phased out in September 2022 (Shlyuger, 2022), which may lose it some users.

### *Altmetrics*

Except for Mendeley and Twitter, the results above are the first reported altmetric correlations with research quality scores and so cannot be compared with prior research. Mendeley correlations are discussed below and Twitter here. Compared to the REF2014 results from 2008 (HEFCE, 2015), the current results are 7-10 years newer and are more robust due to taking a median of several years rather than a single year. Even accounting for this, the Twitter correlations above are surprisingly much stronger than the REF2014 correlations (Tables A54 of: HEFCE, 2015). For 2008, the highest Twitter correlation was 0.23 for Art and Design: History, Practice and Theory, the second highest was 0.17 for Public Health, Health Services and Primary Care, and the remaining correlations were below 0.15, with an average of 0.06. This is only a third of the average correlation above (0.18). Thus, either Altmetric.com's data collection has become more systematic since 2008 or Twitter has changed or matured as a scholarly communication platform. The former seems likely because Altmetric.com was founded in 2011 so its Twitter data for 2008 may well have been incomplete (Thelwall et al., 2013).

The relatively high correlations for health-related fields may reflect widespread public interest in potentially impactful medical research (Mohammadi et al., 2018). This increases the amount of altmetric data but also suggests that the public tends to be interested in higher quality research to some extent. This is despite public interest in health research being very topic driven, for example with particular concern for cancer and especially breast cancer (Lewison et al., 2008).

### *Mendeley readers vs. Scopus citations*

The comparison between Mendeley reader counts and Scopus citation counts above contrasts sharply with the data available from REF2014 (Tables A39 of: HEFCE, 2015). For the 2008 REF2014 data, the Mendeley correlations were overall 52% of the strength of the Scopus citation correlations, with Mendeley being stronger in only 5 out of 36 cases. A likely partial cause of this is Altmetric collecting Mendeley data more systematically now, so its data more closely reflects Mendeley readers. In addition the first Mendeley results above (Figure 1) use comprehensive data from the Mendeley API. Altmetric previously harvested data from Mendeley for articles that it had registered through other altmetrics, so would have missed some results (Thelwall et al., 2013). Mendeley was launched at the end of 2007 (Henning & Reichelt, 2008) and needed some time to generate a substantial userbase, but its 2008 data nevertheless seems to be as substantial as its later data (Thelwall & Sud, 2016). Thus, probably because of incomplete early Altmetric Mendeley data, the HEFCE analysis seems to have underestimated the value of Mendeley as a research quality indicator. The current results suggest that in most fields outside the arts and humanities it is similar in strength to citation counts in this role, although usually a little weaker.

An exception to the above conclusion is that a previous study claimed that Mendeley readers were as useful in the arts and humanities as elsewhere (Thelwall, 2019). The above results suggest that this is false because Mendeley is of little use in the arts and humanities as a quality indicator. It is even substantially less useful than citations, which are themselves very weak research quality indicators.

*Field normalisation*

The comparison between raw Scopus citation counts and field normalised NLCS versions above echo the data available from REF2014, although this was not analysed in the report (Tables A3, A8 of: HEFCE, 2015). For REF2014, Scopus's Field Weighted Citation Impact (FWCI), which is similar to the NLCS above except without the log transformation component (Scopus, 2020), had a stronger correlation with REF2014 final scores for 2008 articles than did raw citation counts in only a third of UoAs (12 out of 36). Thus, similar results for two different field normalised datasets suggest that field normalisation of citation counts is not helpful when comparing articles within the same field, at least when using Scopus narrow fields for normalisation.

# Conclusions

In answer to the first research question, the Altmetric.com altmetrics that are most useful as indicators of research quality are, in descending order, Mendeley readers, Tweeters, Facebook Walls, News, Blogs, Wikipedia, Reddit and Research Highlights. Of these Mendeley is close to Scopus citations in power as a research quality indicator, and Tweeters is clearly the best of the social web indicators. The last three only have minor value. The evidence is the strongest yet for Mendeley and Twitter and is the first of its kind for the others. The results support the continued use of altmetrics as attention indicators by publishers even though the evidence for some is weak. They particularly strengthen the case for the value of Twitter for altmetrics. Doubt had been previously cast on it due to Twitter's use for publicity and spam, but the current results suggest that these uses have either declined, been filtered out by Altmetric.com, or naturally align with the quality of articles. In terms of field differences, altmetrics have the most value in health fields, and the physical sciences and the least value in the arts and humanities.

In answer to the second research question, the proprietary but well known Altmetric Attention Score seems to be an effective quality indicator even though it is designed to reflect attention rather than research quality or impact. This result has the practical implication that it is a reasonable value for researchers to consult informally, especially in the health and physical sciences.

For the third research question, none of the altmetrics are as effective as citation counts as research quality indicators, despite Mendeley being a close second. It is reasonable to continue to use Mendeley readers as a substitute for citations as an early impact or quality indicator when the citation window is too narrow for citations, however.

Finally, an accidental by-product of this research was the unexpected finding that field normalising citations using Scopus narrow fields reduces their value as research quality indicators, presumably due to problems with the field classifications used. Thus, research evaluators should consider avoiding field normalisation when a set of articles to be evaluated are mainly from a single broad field. Alternatively, a more consistent field categorisation scheme might be used (Klavans & Boyack, 2017), if available.

# Acknowledgements

This study was funded by Research England, Scottish Funding Council, Higher Education Funding Council for Wales, and Department for the Economy, Northern Ireland. The content is solely the responsibility of the authors and does not necessarily represent the official views of the funders.